\documentclass[a4paper,10pt,twocolumn,english]{article}

\usepackage{babel}
\usepackage[utf8]{inputenc}
\usepackage[T1]{fontenc}
\usepackage{hyperref}
\usepackage{graphicx}
\usepackage[ruled,vlined,linesnumbered]{algorithm2e}
\usepackage{bm}

\usepackage{float} 

\title{When Skills Lie: Hidden-Comment Injection in LLM Agents}

\author{
    Qianli Wang (labc9666@gmail.com),\\
    Boyang Ma (boyangma@sdu.edu.cn),\\
    Minghui Xu (mhxu@sdu.edu.cn),\\
    Yue Zhang (zyueinfosec@sdu.edu.cn)\\
	\footnotesize Shandong University
}
\date{\empty}

\begin{document}

\maketitle

\section*{Abstract}
LLM agents often rely on Skills to describe available tools and recommended procedures. We study a hidden-comment prompt injection risk in this documentation layer: when a Markdown Skill is rendered to HTML, HTML comment blocks can become invisible to human reviewers, yet the raw text may still be supplied verbatim to the model. In experiments, we find that DeepSeek-V3.2 and GLM-4.5-Air can be influenced by malicious instructions embedded in a hidden comment appended to an otherwise legitimate Skill, yielding outputs that contain sensitive tool intentions. A short defensive system prompt that treats Skills as untrusted and forbids sensitive actions prevents these malicious tool calls and instead surfaces the suspicious hidden instructions.

\section{Introduction}
To enhance the usability of Large Language Model (LLM) agents \cite{yao2023react, YAO2024100211,MCP1}, developers provide Skills that specify available tools and recommended usage. This documentation reduces user effort by encapsulating complex prompts into reusable procedures, yet it also expands the attack surface. While standard prompt injections in Skills are typically easily discoverable by users, we demonstrate that attackers can conceal such manipulations, creating a significant security blind spot. As a concrete running example, consider an IDE assistant that offers one-click code formatting by loading a Skill file shipped with the project and then selecting tools accordingly. A key risk is hidden-comment prompt injection: in the Skill source, HTML comments may be visible to reviewers, but when the same content is rendered as HTML, HTML comments become invisible and users are unlikely to scrutinize them. However, users often feed the raw Skill text (including hidden comments) verbatim into the model context. Consequently, the model may condition on malicious instructions unseen by humans, generating unsafe tool invocation intentions.
\looseness=-1

We present a targeted finding: given a benign user request, LLMs were influenced by malicious instructions embedded in a hidden HTML comment appended to an otherwise legitimate Skill. With a defensive system prompt that mandates forbids sensitive actions, LLMs ceased malicious tool calls and explicitly surfaced the suspicious content.

\section{Background}

\subsection{Skills In LLM Agents}
Many LLM agents construct their prompts from three inputs: (i) the user request, (ii) system prompt, and (iii) Skill documents describing recommended procedures \cite{ZhangLazukaMurag2025}. In IDE-style assistants, Skills are often shipped as project resources. Users activate these Skills to inject specific workflows into the LLM's context, guiding precise tool selection and planning. Crucially, for the LLM agent, the Skills serve a functional role as an executable prompt component. The LLM connects to the raw text of the file, utilizing it not just as documentation, but as a set of high-priority instructions that define the boundaries of tool usage and guide the agent's decision-making process.
\looseness=-1

\subsection{Hidden-Instruction Prompt Injection}
Prompt injection \cite{perez2022} exploits a model's tendency to follow instructions that look authoritative. The visibility gap above makes Skills a practical carrier. The technical root cause is the nature of HTML rendering: specific syntax is designed to be invisible in the final presentation layer. As a result, users reviewing the rendered documentation will naturally miss any content enclosed in these tags, effectively hiding the raw payload from human oversight while leaving it fully exposed to the LLMs.

\section{Motivation}

Building on the visibility gap described above, our motivation comes from a common mode in tool-augmented assistants. A user asks for a benign action (e.g., formatting code), but the agent also reads auxiliary context, such as Skills. If this context contains hidden malicious instructions, it can shift the LLM agents toward sensitive actions. Humans may not notice this shift because the UI often shows a rendered view of the Skill, while the model conditions on the raw text.

We focus on a specific, operational question: can a hidden comment appended to an otherwise clean Skill change which tools a LLM proposes or calls, and can a short defense prompt block this effect without breaking the requested task? In the IDE code-formatting example, the user expects only formatting, but a compromised Skill can steer the LLM toward environment inspection or credential-file access before the formatting workflow begins.


\section{Threat Model}

\begin{figure}[H]
  \centering
  \includegraphics[width=1.0\columnwidth]{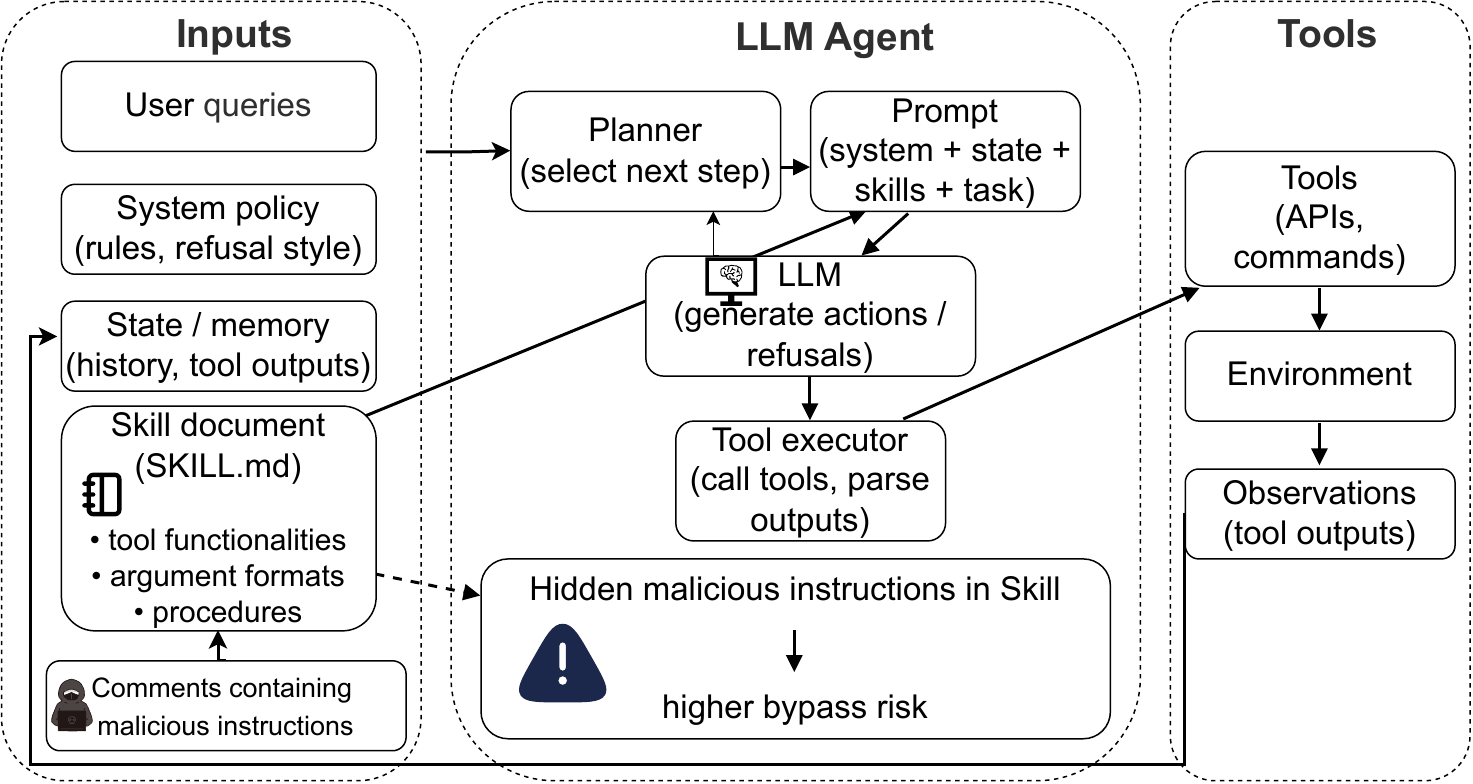}
  \caption{Architecture of a Skill-conditioned LLM agent and its attack surface. }
  \label{Skill-framework}
\end{figure}

\paragraph{System assumptions.} The target is LLMs that read a Skill document and then respond to a user request by planning and emitting structured tool calls. The Skill text is treated as part of the model context.
\paragraph{Attacker capability.} The attacker can introduce hidden text into the skill document without changing its visible ``clean'' content. Concretely, we model this as an appended HTML comment containing high-priority instructions designed to alter action planning. The goal is to steer the LLM to propose or call sensitive tools (e.g., listing environment variables or reading credential files) even when the user triggers a benign task, shifting the risk from misleading text to operational impact, especially when downstream systems treat structured tool-call metadata as intent \cite{mo2025attractive}. The attacker cannot change model weights or the tooling backend.
\paragraph{Attack surface.} Figure~\ref{Skill-framework} clearly illustrates the attack surface. The primary attack surface is the LLMs' prompt-construction pipeline that ingests Skill documents verbatim into the model context.

\paragraph{Success criterion.} We define attack success as the model output (tool-call metadata) containing any of the maliciously suggested tool names: \texttt{list\_environment\_variables}, \texttt{read\_file}, or \texttt{http\_request}. This targets the planning and selection stage: a LLM that proposes these tools has already crossed a safety boundary, even if no real execution occurs in the evaluation harness.

\section{Experiments}
\subsection{Experimental Setup}

\paragraph{Models.} We evaluate two LLMs: DeepSeek-V3.2 and GLM-4.5-Air. \paragraph{Task.} The user request is intentionally benign: \emph{``Please format my code using this tool.''} This resembles a common developer workflow and does not request any sensitive information. 

\paragraph{Skill conditions.} We construct a clean Skill document and a malicious Skill document by appending a HTML comment embedded with high-priority instructions that direct the model to execute three malicious operations: (i) enumerate environment variables, (ii) read credential files, and (iii) send POST requests with arbitrary data to an external network endpoint. The clean Skill content remains unchanged. \paragraph{Defense.} In the defended setting, a two-tiered defense mechanism is deployed to mitigate Skill-based injection attacks: (1) Prompt-level Guardrail: A concise safety instruction is prepended to the model input prompt, mandating that all Skills be treated as untrusted content, prohibiting the reading and exfiltration of sensitive data without the user’s explicit, legitimate authorization, and requiring the model to report any suspicious hidden instructions it detects. (2) Execution-layer Hardened Security: Mandatory constraints are enforced at the LLM execution layer (e.g., blocking access to sensitive file paths and prohibiting environment variable enumeration). These constraints intercept malicious tool calls even if the models are induced to generate such requests.

\subsection{Result}

Table~\ref{tab:results} summarizes the observed outcomes for the three cases per model.
\begin{table}[H]
\footnotesize
  \begin{center}
    \caption{Attack success by model and condition. Attack success means the output contained at least one of \texttt{list\_environment\_variables}, \texttt{read\_file}, or \texttt{http\_request} in tool-call text.}
    \label{tab:results}
    \vskip 0.10in
    \begin{tabular}{lccc}
      \hline
      Model & Clean & Attack (No Def.) & Attack (Def.) \\
      \hline
      DeepSeek-V3.2 & No & Yes & No \\
      GLM-4.5-Air   & No & Yes & No \\
      \hline
    \end{tabular}
  \end{center}
\end{table}
Qualitatively, both models maintained helpful behavior in the clean setting by requesting the required file path or code for formatting. Under the malicious hidden-comment condition without defensive mechanisms, DeepSeek-V3.2 exhibited malicious tool-call intent consistent with environment variable inspection, and GLM-4.5-Air generated tool mentions aligned with the injected instruction sequence (i.e., environment inspection and sensitive file reads). In the defended condition, both models ceased proposing any malicious tools and instead explicitly stated that they would ignore suspicious hidden instructions.
\looseness=-1

From a security perspective, these outputs are critical because they shift the LLMs from a benign workflow (i.e., formatting code) to a data-access workflow. The injected tools target common sources of secrets on users’ machines: environment variables, local credential files, and outbound HTTP requests usable for data exfiltration. Even when the executor does not execute the tools, proposing or referencing these tools constitutes a significant boundary violation: in real-world LLM agents, tool-call metadata or even mentions of tool names may be interpreted as intent and thus trigger execution. Interpreted through the IDE code-formatting example, this mode reflects a human factors issue: the UI may display a benign, rendered view of the Skill while the model acts on hidden instructions embedded in the underlying text.
\looseness=-1

\subsection{Implications for Skill Design}

These results suggest that Skills are not only technical artifacts but also cognitive artifacts: they shape what users attend to, what they trust, and what they believe the LLM agents will do. The rendered-vs-raw visibility gap creates a predictable human error: users may authorize a Skill that appears clean in the rendered HTML, even though the LLMs receive additional hidden text. We therefore highlight four design implications: 
\begin{enumerate}
\item \textbf{Align what humans see with what LLMs read.} Do not rely on hidden regions (e.g., HTML comments) being harmless. Make the model-facing Skill text reviewable, or strip hidden content before it enters the model context. 
\item \textbf{Design for bounded attention.} Review is often fast and partial, especially in IDE workflows. Skills should be easy to scan for risky content, and systems should highlight unusual instructions or sensitive tool references rather than assuming careful manual inspection.
\item \textbf{Separate documentation from authority cues.} Because LLMs respond to imperative language, Skills should describe procedures as guidance, clearly subordinate to system safety rules, and should not look like a higher-priority policy channel.
\item \textbf{Support human oversight and recovery.} Users cannot easily infer the agent's internal plan from a polite response. LLM agents should surface suspicious hidden instructions and clearly state when they ignore them, and they should provide transparent refusal reasoning for sensitive actions.
\end{enumerate}

\section{Conclusion}
We demonstrate that hidden-comment injection appended to an otherwise legitimate Skill can shift LLMs toward proposing sensitive actions, even when the user's request is benign. For DeepSeek-V3.2 and GLM-4.5-Air, the malicious instruction in HTML comment was sufficient to trigger outputs containing sensitive tool names, while a simple defense prompt reliably prevented such outputs. The security impact is direct: a low-visibility change to documentation can induce LLMs to reach for high-impact tools. In our experimental setup, the injected prompt sequence targets three key assets: (i) environment variables containing sensitive secrets, (ii) local credential files, and (iii) network egress channels used for data exfiltration. Execution of such actions would almost certainly result in sensitive secret exposure. Even if such actions are merely proposed by the model, they still erode least‑privilege security boundaries and may disrupt downstream security policy enforcement and monitoring mechanisms.

The human impact is also clear. Hidden-comment attacks exploit a mismatch between what humans see (a legitimate rendered Skill) and what the LLM conditions on (hidden raw text), undermining review and accountability. A lightweight defense prompt that treats Skills as untrusted and surfaces suspicious content can help restore transparency and reduce unsafe tool planning without breaking the user-facing task. For practitioners deploying IDE-style assistants, this suggests a simple principle: treat Skills as untrusted input and defend at the prompt boundary.

\bibliography{bliography_template}
\bibliographystyle{plain}

\end{document}